# Localized States in the Chaotic Ce Atom

Mario Feingold

*Dept. of Physics, Ben-Gurion University,*

*Beer-Sheva 84105, Israel*

and

Oreste Piro

*Departament de Física, Universitat de les Illes Balears,*

*07071 Palma de Mallorca, Spain*

ABSTRACT: We show that the energy averaged entropy localization length for the eigenstates of the Ce atom is well approximated by the prediction of the Wigner ensemble.

In Ref. 1, predictions of the Wigner ensemble[2] of banded random matrices are compared with the results obtained in a model of the Ce atom. In particular, it is shown that, in the basis of the single electron orbitals arranged in increasing order of the corresponding energies, the eigenstates of Ce are localized (see Fig. 12 of Ref. 1). This finding, although in agreement with recent predictions,[3,4,5,6,7,8] contradicts the widely accepted expectation that the eigenstates of time-independent strongly chaotic systems are generically extended[9] and therefore, it is important.

On the quantitative side, the comparison between the average shape of the Wigner ensemble squared eingenvectors, $< |v_{ij}|^2 > = w(i-j)$, and the eigenstates of Ce leads to some agreement the extent of which is however unclear (see Figs. 13 and 20 of Ref. 1). This comparison is hampered by the fact that $w$ is only implicitly known as the solution of a complicated nonlinear integral equation.[2] It can be explicitly obtained only for $|i-j| < b$ and $|i-j| \gg b$ where $b$ is the band width of the random matrices.[10] Accordingly, in Ref. 1, $\rho_w(u) \equiv w(\frac{i-j}{b})$ was separatedly compared in the two regimes with the corresponding Ce eigenfunctions. In the small $u$ range, $\rho_w$ is found to be close to the Lorentzian predicted by the Wigner ensemble but having lower tails. Furthermore, in the large $u$ regime, where the better agreement is claimed, $\rho_w$ is compared with the individual eigenvectors of Ce rather than with their average. Since in the Ce atom the parameters corresponding to those of the random matrix ensemble, $b$ and $\alpha$, are varying with energy, it is unjustified to expect such detailed agreement between the two. In what follows, we propose that instead a more global feature of the eigenvectors, e.g. the *average* entropy localization length,[11] $L$, should be used as a measure of the agreement between the properties of the Ce atom and the predictions of the Wigner ensemble. Moreover, we show that the values of $L$ for the two models are extremely close to each other.

We use the data of Ref. 1 to calculate the energy averaged values of $b$, $\alpha$ and $L$, $b_{\text{Ce}}$,

$\alpha_{\text{Ce}}$ and $L_{\text{Ce}}$, respectively (see Figs. 3, 14, 21 and 22 of Ref. 1). One obtains $b_{\text{Ce}} = 78.7$, $\alpha_{\text{Ce}} = 0.0266$ and $L_{\text{Ce}} = 131.1$ for the $J^\pi = 4^-$ states (odd). Moreover, $N_{\text{Ce}}$, the size of the single electron orbitals basis used in the calculations of the Ce atom model, is 260. On the other hand, the numerical computation of $L$ for 100 random matrices with the same parameters, gives $L_W = 137.4 \pm 0.8$ for $b = 78$ and $L_W = 138 \pm 1$ for $b = 79$. We incline to regard this result as being in very good agreement with that for the Ce atom. However, one should estimate the size of the error of $L_{\text{Ce}}$ in order to strengthen this statement. This error is both due to our approximate handling of the data of Ref. 1 and the energy variation of the parameters in the Ce atom which is not accounted for by the Wigner ensemble. It is expected to be of the order of a few percent.

It is worthwhile to point out the relation of these results to the recent studies on the behavior of $L_W$. It was shown[3,4,5,6,7,8] that in the scaling regime, namely, for $z \equiv \alpha^{-1} b^{-1/2} \gtrsim 0.3$, $L_W = b^2 f(x, y)$ where $x = b^2/N$ and $y = \alpha b^{3/2}$. In the Ce atom $z = 0.42$, not too far from the edge of the scaling regime. Moreover, $x_{\text{Ce}} = 23.8$ and $y_{\text{Ce}} = 185.9$, which according to Ref. 6 corresponds to a regime with Anderson-like behavior and $L_W = 118.7$. Since the values of $x$ and $y$ for the Ce atom are much larger than the largest ones considered in Ref. 6, the inaccuracy in the predicted value of $L_W$ is not surprising. As fas as the spacings distribution, $P(s)$, is concerned the prediction of the Wigner ensemble is that it should follow the Wigner distribution (see Fig. 4 of Ref. 6). This is indeed observed in the Ce atom (see Fig. 2 of Ref. 1) and in all the other strongly chaotic models studied up to now. It is presently not clear however, whether or not strongly chaotic systems can have values of $x$ and $y$ for which the $P(s)$ of the corresponding Wigner ensemble deviates from the Wigner distribution.[7]

Finally, we suggest that a simpler model than the Ce atom should be studied in order to verify the predictions of the Wigner ensemble. In particular, the classical dynamics for such model should be studied to make sure that it is fully chaotic.

We would like to thank G. Gribakin for useful discussions. MF acknowledges the support of an Allon fellowship and OP that of Dirección General de Investigación Científica y Técnica, contract number PB92-0046-c02-02. This work was done during a visit of MF at the Universitat de les Illes Balears. He thanks the Departament de Física for its hospitality.